\documentclass[aps,showpacs, prl,twocolumn,superscriptaddress]{revtex4}%
\usepackage{epsfig,dsfont,amssymb,amsmath,amsthm,amsfonts,amsbsy,mathrsfs}
\usepackage{graphicx}
\usepackage{amsmath}
\usepackage{amssymb}
\begin{document}
\title{Structure, stability and electronic properties of tricycle type graphane}
\author{Chaoyu He}
\affiliation{Institute for Quantum Engineering and Micro-Nano Energy
Technology, Xiangtan University, Xiangtan 411105, China}
\author{L. Z. Sun}
\email{lzsun@xtu.edu.cn} \affiliation{Institute for Quantum
Engineering and Micro-Nano Energy Technology, Xiangtan University,
Xiangtan 411105, China}
\author{C. X. Zhang}
\affiliation{Institute for Quantum Engineering and Micro-Nano Energy
Technology, Xiangtan University, Xiangtan 411105, China}
\author{N. Jiao}
\affiliation{Institute for Quantum Engineering and Micro-Nano Energy
Technology, Xiangtan University, Xiangtan 411105, China}
\author{K. W. Zhang}
\affiliation{Institute for Quantum Engineering and Micro-Nano Energy
Technology, Xiangtan University, Xiangtan 411105, China}
\author{Jianxin Zhong}
\email{zhong.xtu@gmail.com}\affiliation{Institute for Quantum
Engineering and Micro-Nano Energy Technology, Xiangtan University,
Xiangtan 411105, China}
\date{\today}

\begin{abstract}
We propose a new allotrope of graphane, named as tricycle graphane,
with a 4up/2down UUUDUD hydrogenation in each hexagonal carbon ring,
which is different from previously proposed allotropes with UUDUUD
(boat-1) and UUUUDD (boat-2) types of hydrogenation. Its stability
and electronic structures are systematically studied using
first-principles method. We find that the tricycle graphane is a
stable phase in between the previously proposed chair and stirrup
allotropes. Its electronic properties are very similar to those of
chair, stirrup, boat-1 , boat-2, and twist-boat allotropes. The
negative Gibbs free energy of tricycle graphane is -91 meV/atom,
which very close to that of the most stable chair one (-103
meV/atom). Thus, this new two-dimensional hydrocarbon may be
produced in the process of graphene hydrogenation with a relative
high probability compared to other conformers. \\
\end{abstract}
\maketitle
\section{Introduction}
\begin{figure*}
\centering
\includegraphics[width=5.5in]{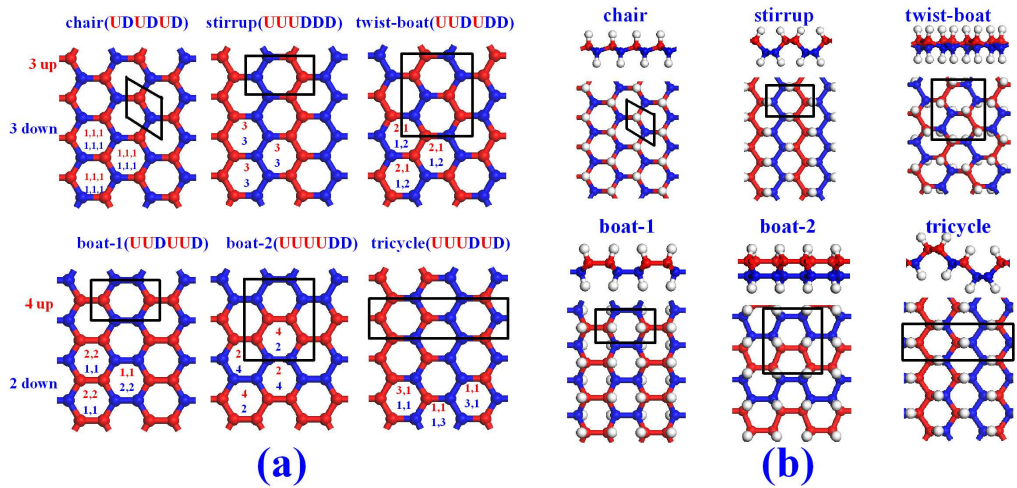}\\
\caption{(a) Schematic diagram of six possible configurations of
hydrogenated graphene with equivalent hexagonal hydrocarbon rings.
(b) Crystal structures (side and top views) of graphane with chair,
stirrup, twist-boat, boat-1, boat-2 and tricycle configurations,
respectively. In figures, the red and blue balls correspond to
carbon atoms with up and down hydrogenation, respectively, and the
white balls are hydrogen atoms.}\label{fig1}
\end{figure*}

\indent  Graphene, a single layer of graphite, is the thinnest
two-dimensional (2D) carbon material consisting of a monolayer
carbon atoms in a honeycomb lattice. This novel material was
experimentally discovered in 2004 \cite{1} and has given rise to
enormous scientific and technological impacts in relevant areas of
physics, chemistry and materials sciences. It is considered as a
revolutionary material for future generation of high-speed
electronics, sensors, transparent electrodes due to its unusual
electronic, optical and magnetic properties. For the purpose of
modifying graphene for applications in future nano-electronics, many
modified graphene-based materials, such as the graphene nanoribbons
(GNRs) \cite{2, 3, 4} and hydrogenated-graphene (graphane) \cite{5,
6, 7, 8, 9, 10, 11, 12}, have been proposed and widely investigated. \\
\indent Graphane, a 2D hydrocarbon material, was suggested first by
Sluiter and Kawazoe \cite{5} and synthesized \cite{7} in 2009
through exposure of a single-layer graphene to a hydrogen plasma.
Since then, many new configurations with low energies for this 2D
hydrocarbon were proposed. The most stable configuration of graphane
is named as "chair" \cite{5, 6} with the UDUDUD hydrogenation in
each hexagonal carbon ring as shown in Fig. \ref{fig1} (a) (here we
adopt the nomenclature in reference [11]). The second stable
configuration named as "stirrup" \cite{5, 9, 10} with the UUUDDD
hydrogenation in each carbon ring is shown in Fig. \ref{fig1} (a) ,
whose energy is about 28 meV/atom larger than that of the chair one.
From the point of view of stability, the following configurations
for graphane allotropes are boat-1 \cite{5, 6, 9} with the UUDDUU
hydrogenation, boat-2 \cite{9, 11} with the UUUUDD hydrogenation,
twist-boat \cite{12} with the UUDUDD hydrogenation and other
configurations with relatively high energies reported in reference
[8, 11]. The hydrogenation patterns of all above graphane allotropes
can be divided in to two groups: 3up/3down and 4up/2down.
Interestingly, we notice that the hexagonal hydrocarbon rings in the
most stable five graphane allotropes, namely chair, stirrup, boat-1,
boat-2 and twist-boat, are equivalent. As mentioned in reference
[11], there are many situations to hydrogenate a single six-carbon
ring, such as 6U: UUUUUU; 5U: UUUUUD; 4U: UUUUDD, UUUDUD, and
UUDUUD; 3U: UDUDUD, UUDUDD, and UUUDDD, 2U= 4U, etc. In present
work, with the restrictive condition of keeping the hexagonal
hydrocarbon rings equivalent in the systems, we propose a tricycle
graphane allotrope where each hexagonal hydrocarbon rings with the
same UUUDUD hydrogenation are equivalent. This new grahane holds
remarkable stability comparable to the most stable chair one,
becoming the second stable graphane allotrope.\\
\begin{table*}
  \centering
  \caption{Structure information: space group (SG), lattice constant (LC), inequivalent atom positions (positions) for H and C atoms,
  C-H bond length (L$_{CH}$), and C-C bond length (L$_{CC}$)) for the six fundamental allotropes of graphane}\label{tabI}
\begin{tabular}{c c c c c}
\hline \hline
System &SG and LC &positions  &L$_{CH}$ and L$_{CC}$\\
\hline
chair      &P-3m1 (164) ,                  &H:(0.3333, 0.6667, 0.5893)  &C-H: 1.110\\
UDUDUD     &a=b=2.504; c=15.0             &C:(0.3333, 0.6667, 0.5153)  &C-C: 1.537\\
tricycle   &Pbcm (57)                     &H1:(0.4328, 0.1235, 0.2500) &C$_1$-H$_1$: 1.108\\
UUUDUD     &a=15; b=7.681; c=2.544        &C1:(0.4981, 0.0563, 0.2500) &C$_1$-C$_1$: 1.539; C$_1$-C$_2$: 1.541\\
           &                              &H2:(0.6364, 0.1190, 0.2500) &C$_2$-H$_2$: 1.109\\
           &                              &C2:(0.5731, 0.1934, 0.2500) &C$_2$-C$_2$: 1.540; C$_2$-C$_1$: 1.541\\
stirrup    &Pmna (53)                     &H:(0.0000, 0.3983, 0.5085)  &C-H: 1.105\\
UUUDDD     &a=2.549; b=15.0; c=3.828      &C:(0.0000, 0.3639, 0.4620)  &C-C: 1.544\\
boat-1     &pmmn (59)                     &H:(0.5000, 0.2562, 0.5922)  &C-H: 1.105\\
UUDDUU     &a=2.529; b=4.309; c=15.0      &C:(0.5000, 0.1822, 0.5216)  &C-C: 1.537; 1.570\\
boat-2     &Pbcm (57)                     &H:(0.3987, 0.4932, 0.5036)  &C-H: 1.103\\
UUUUDD     &a=15.0; b=4.585; c=4.328      &C:(0.4622, 0.5939, 0.4317)  &C-C: 1.542; 1.548; 1.573\\
twist-boat &Pcca (54)                     &H:(0.1215, 0.4079, 0.5609)  &C-H: 1.106\\
UUDUDD     &a=4.417; b=15.0; c=4.987      &C:(0.0904, 0.4788, 0.6154)  &C-C: 1.542; 1.548; 1.562\\
\hline \hline
\end{tabular}
\end{table*}
\section{Models and Methods}
\indent To hydrogenate a six-carbon ring, there are finite
situations (6U: UUUUUU; 5U: UUUUUD; 4U: UUUUDD, UUUDUD, UUDUUD; 3U:
UDUDUD, UUDUDD, UUUDDD; 2U=4U, etc) according to mathematical
permutation and combination principle. However, there are enormous
situations for hydrogenated graphene because there are considerable
patterns of the border between hexagonal carbon rings. Moreover,
each carbon atom in graphene belongs to three neighboring hexagonal
carbon rings and its hydrogenated configuration (up/down) will be
counted into the hydrogenation situations of its three neighboring
six-carbon rings, making the situation very complicated. However, we
realize that the hexagonal hydrocarbon rings of the five most stable
graphane allotropes, including chair, stirrup, boat-1, boat-2 and
twist-boat configurations, are equivalent in each system, as shown
in Fig.\ref{fig1}. Under the restrictive condition of keeping the
equivalence of the hexagonal hydrocarbon rings in the systems,
through re-examining the systems with 4U and 3U hydrogenations
(here, 4U and 3U are identical to 2U and 3U systems, and 6U and 5U
are abandoned due to their instability), we find a new hydrogenated
graphene with 4up/2down configuration satisfying the equivalent
requirement as in the five most stable graphane allotropes. Such
graphane allotrope is named as tricycle allotrope. The hydrogenation
patterns of all the five most stable graphane allotropes proposed
previously and the one proposed in our present work are indicated in
Fig. \ref{fig1} (a). They can be divided in to two groups, namely,
3up/3down and
4up/2down. Their corresponding crystal structures including side and top views are shown in Fig. \ref{fig1} (b).\\
\begin{table}
  \centering
  \caption{The calculated Cohesive energy (E$_{coh}$: eV/atom), Gibbs free energy ($\delta$G: meV/atom),
  band gap (Eg: eV), layer thickness(LT: {\AA}) and surface work function (SWF: eV) for the six graphane allotropes.}\label{tabII}
\begin{tabular}{c c c c c c c}
\hline \hline
Item             &chair &tricycle &stirrup &boat-1 &boat-2 &twist-boat\\
\hline
E$_{coh}$        &-5.222 &-5.210   &-5.194 &-5.171     &-5.155 &-5.147\\
$\delta$G        &-103   &-91.2    &-74.3  &-51.7      &-35.6  &-27.8\\
Band gap         &3.491   &3.446    &3.340   &3.374    &3.412   &3.529\\
LT               &2.679   &4.093    &3.053   &2.758    &3.039   &2.761\\
SWF              &3.916   &4.528    &4.407   &4.416    &4.457   &4.543\\
 \hline \hline
\end{tabular}
\end{table}
\indent The structures, relative stability and electronic properties
of these six fundamental 2D graphane allotropes are systematically
investigated using first-principles methods in the framework of
density functional theory (DFT). All calculations are performed
within the general gradient approximation (GGA) \cite{13} as
implemented in Vienna ab initio simulation package (VASP) \cite{14,
15}. The interactions between nucleus and the valence electrons are
described by the projector augmented wave (PAW) method \cite{16,
17}. A plane-wave basis with a cutoff energy of 500 eV is used to
expand the wave functions. The Brillouin Zone (BZ) sample meshes are
set to be dense enough (less than 0.21/{\AA}) to ensure the accuracy
of our calculations. Crystal lattices and atom positions of all
graphane allotropes are fully optimized up to the
residual force on every atom less than 0.01 eV/{\AA} through the conjugate-gradient algorithm.\\
\section{Results and Discussion}
\subsection{Structures}
\indent The optimized structures of chair, stirrup, boat-1, boat-2,
twist-boat and tricycle graphane allotropes are shown in Fig.
\ref{fig1} (b) and their structural information is summarized in
Tab. \ref{tabI}. The lattice constants for chair (P-3m1) and boat-1
(Pmmn) graphane allotropes are (a=2.504 {\AA}, b=2.504 {\AA}, c=15.0
{\AA}) and (a=2.529 {\AA}, b=4.309 {\AA}, c=15.0 {\AA}),
respectively. The lattice constants, H-H bond length and C-C bond
length for chair and boat-1 graphane allotropes are in good
agreement with previous report\cite{6}. Such consistency confirms
the correctness of our calculations. The lattice constants for
boat-2 (Pbcm), stirrup (Pmna) and twist-boat (Pcca) graphane
allotropes are (a=15.0 {\AA}, b=4.585 {\AA}, c=4.328 {\AA}),
(a=2.549 {\AA}, b=15.0 {\AA}, c=3.828 {\AA}), and (a=4.585 {\AA},
b=15.0 {\AA}, c=4.328 {\AA}), respectively. All these five graphane
allotropes have only one inequivalent CH pair in their crystal cell.
The inequivalent H and C atom positions can be found in Tab.
\ref{tabI} and elsewhere \cite{6, 10, 12}. The tricycle graphane
allotrope proposed in present work contains two pairs of
inequivalent CH. However, similar to the previously proposed
allotropes, it contains only one inequivalent six-carbon ring.
Tricycle graphane belongs to Pbcm space group and its lattice
constants are a=15.0 {\AA}, b=7.681 {\AA} and c=2.544 {\AA}. Four
inequivalent atoms in the unit cell of tricycle graphane locate at
positions of (0.4328, 0.1235, 0.2500), (0.4981, 0.0563, 0.2500),
(0.6364, 0.1190, 0.2500) and (0.5731, 0.1934, 0.2500) for H1, C1, H2
and C2, respectively. The calculated bond lengths are 1.108 {\AA},
1.539 {\AA}, 1.5109 {\AA}, 1.540 {\AA} and 1.541 {\AA} for
C$_1$-H$_1$, C$_1$-C$_1$, C$_2$-H$_2$, C$_2$-C$_2$ and C$_1$-C$_2$,
respectively. Its bond lengths are very close to those in chair,
boat-1, boat-2, stirrup and twist-boat allotropes. The layer
thicknesses of these six allotropes are summarized in Tab.
\ref{tabII}. We can see that the most thinnest is chair allotrope
with a thickness of 2.6779 {\AA} and the thickest tricycle allotrope
holds thickness of 4.093 {\AA}. The large thickness of tricycle
graphane provides a possible way to identify its existence in
experiments. Surface work function provides another way to identify
these potential graphane allotropes. From Tab. \ref{tabII}, we can
see that the work function of the chair graphane is 3.916 eV, which
is the smallest. The work functions for other four graphane
allotropes of tricycle, stirrup, boat-1, boat-2 and twist-boat are
4.528 eV, 4.407 eV, 4.416 eV, 4.457 eV and 4.543 eV, respectively.
one can identify these allotropes through the values of their
surface work functions in field emission experiments.\\
\subsection{Stability}
\indent The relative stability of these allotropes can be evaluated
through comparing their cohesive energy per atom (E$_{coh}$). Low
energy usually means high probability to be discovered in the same
experimental condition. To evaluate the thermodynamic stability of
these new allotropes of graphane, the Gibbs free energy ($\delta$G)
is calculated according to
$\delta$G=E$_{coh}$-$x_H$$\mu_H$-$x_C$$\mu_C$, where $E_{tot}$ is
the cohesive energy per atom of the graphanes with different
compositions, x${_i}$ means the molar fraction of atom i (i=H, C) in
the structure with x${_H}$+x${_C}$=1, and $\mu_i$ is the chemical
potential of each constituent atom. The $\mu_H$ is chosen as the
binding energy per atom of the H${_2}$ molecular and $\mu_C$ is the
cohesive energy per atom of a single graphene. The calculated
cohesive energies and Gibbs free energies for all graphane
allotropes are listed in Tab. \ref{tabII}. We can see that the most
stable one is chair graphane. Its negative $\delta$G of -103
meV/atom indicates high probability to be synthesized from graphene
and H$_2$. The tricycle graphane proposed is the second stable
graphane allotrope with only 12 meV/atom of $\delta$G above the most
stable chair one, indicating that it may be produced in the process
of graphene hydrogenation. It is amazing that this new allotrope
with 4up/2down configuration is more stable than the stirrup one
with 3up/3down configuration, which is contrary to the intuitive
knowledge that balanced up and down hydrogenation configuration is
more stable than that of unbalanced one, (for example,  chair and
stirrup with 3up/3down hydrogenation are more stable than boat-1 and
boat-2 with 4up/2down hydrogenation). Stirrup graphane with UUUDDD
hydrogenation of 3up/3down is now the third stable conformer holding
about 28 meV/atom of $\delta$G larger than that of the chair one.
The value of $\delta$Gs for stirrup, boat-1 and boat-2 are -74
meV/atom, -52 meV/atom and -36 meV/atom, respectively. The negative
$\delta$Gs indicate that all of them may be produced in the process
of graphene hydrogenation. The least stable graphane allotrope
studied in our present work is the twist-boat one with UUDUDD
hydrogenation holding energy about 80 meV/atom larger than that of
the chair one. However, its negative $\delta$G of -28 meV/atom
indicates that it can also be produced through the
hydrogenation process from H$_2$ and graphene.\\
\begin{figure}
\centering
\includegraphics[width=3.5in]{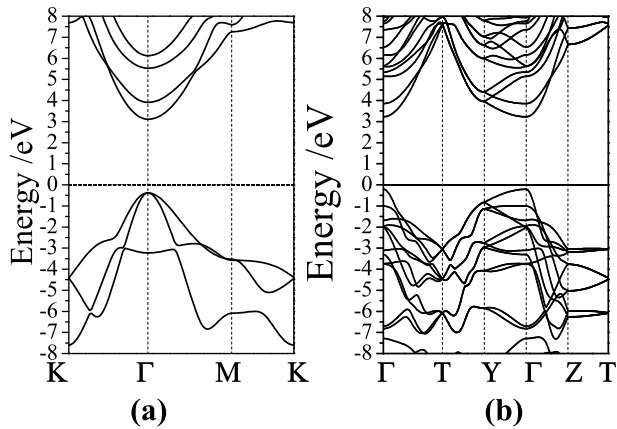}\\
\caption{Band structures of chair graphane (a) and tricycle graphane
(b).}\label{fig2}
\end{figure}
\begin{figure}
\centering
\includegraphics[width=3.50in]{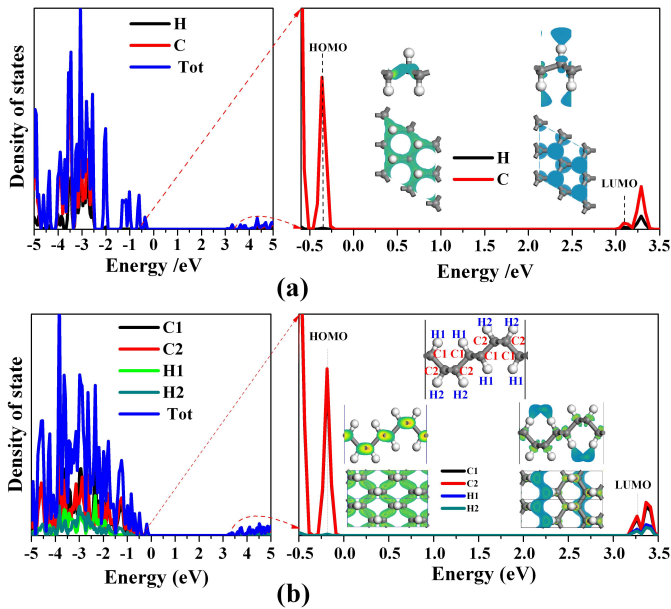}
\caption{Density of states for chair graphane (a) and tricycle
graphane (b). Insets show the charge density distributions of to
HOMO and LUMO states.}\label{fig3}
\end{figure}
\subsection{Electronic properties}
\indent Hydrogenation converts the hybridization of carbon atoms in
graphene from sp$^2$ to sp$^3$, inducing separation between the
valence band and conduction band of graphene. All graphane
allotropes with full coverage of hydrogen are direct-band-gap
semiconductors with band gaps distributing in the range from 3.37 eV
to 3.53 eV. The calculated band structures for the most stable chair
graphane and the second stable tricycle graphane are shown in Fig.
\ref{fig2} (a) and (b), respectively. It is clear that the highest
occupied molecular orbit (HOMO) and the lowest unoccupied molecular
orbit (LUMO) in these two stable hydrocarbon systems are located at
Gama point. The band gaps are 3.491 eV and 3.446 eV for chair and
tricycle graphanes, respectively. In Fig. \ref{fig3}, we show the
local density of state of each inequivalent atom for chair and
tricycle graphanes in (a) and (b), respectively. The right panel of
Fig. \ref{fig3} is the local density of states within the energy
widows of (-0.5 eV, 3.5 eV) and the charge density distributions of
the HOMO and the LUMO states for both systems are also shown in the
figure. From \ref{fig3} (a) we can see that the HOMO states of chair
graphane are mainly from p$_z$ states of carbon atoms and the
corresponding charge densities mainly distribute along the C-C
bonds. Its LUMO states are from both the p$_z$ states of carbon
atoms and the s states of hydrogen atoms and the corresponding
charge densities mainly distribute along the C-H bonds and the weak
H-H bonds. For the tricycle graphane, its LUMO states are mainly
contributed by the p$_z$ states of C$_1$ and C$_2$ atoms and the s
states of H$_1$ and H$_2$ atoms and the corresponding charge
densities mainly distributed along the C$_1$-H$_1$ and C$_2$-H$_2$
bonds and the weak H$_1$-H$_1$ bonds. HOMO states of tricycle
graphane are mainly from the p$_z$ states of carbon atoms as shown
in Fig. \ref{fig3} and the corresponding charge densities mainly
distribute on the C$_1$-C$_1$ bonds and the C$_2$-C$_2$ bonds in the
two inequivalent C$_1$ zigzag-chain and C$_2$ zigzag-chains,
respectively, and are absent on the inter-chains C$_1$-C$_2$ bonds.\\
\section{Conclusion}
\indent  In summary, using first-principle calculations within the
frame work of density functional theory, a tricycle graphane
allotrope was proposed as the sixth fundamental graphane allotropes
with a UUUDUD hydrogenation configuration satisfying the requirement
of keeping each hexagonal hydrocarbon ring equivalent. Our
calculations indicate that tricycle graphane is the second stable
one in the graphane allotropes family proposed so far. Its negative
Gibbs free energy indicates its high probability to be produced in
the process of graphene hydrogenation.\
\section{acknowledgments} This work is supported by the
National Natural Science Foundation of China (Grant Nos. 11074211,
10874143 and 10974166), the National Basic Research Program of China
(2012CB921303), the Cultivation Fund of the Key Scientific and
Technical Innovation Project, the Program for New Century Excellent
Talents in University (Grant No. NCET-10-0169), and the Scientific
Research Fund of Hunan Provincial Education Department (Grant Nos.
10K065, 10A118, 09K033).\\


\begin{thebibliography}{17}
\bibitem{1} Novoselov K S, Geim A K, Morozov S V, Jiang D, Zhang Y, Dubonos S V, Grigorieva I V and Firsov A A. \emph{Science} 306 666 2004.
\bibitem{2} Han M Y, Ozyilmaz B, Zhang Y B and Kim P. \emph{Phys. Rev. Lett} 98 206805 2007.
\bibitem{3} Nakada K, Fujita M, Dresselhaus and Dresselhaus M S. \emph{Phys. Rev. B} 54 17954 1996.
\bibitem{4} Son Y-W, Cohen M L and Louie S G. \emph{Phys. Rev. Lett} 97 216803 2006.
\bibitem{5} Sluiter M H F and Kawazoe. \emph{Phys. Rev. B} 68 085410 2003.
\bibitem{6} Sofo J O, Chaudhari A S and Barber G D. \emph{Phys. Rev. B} 75 153401 2007.
\bibitem{7} Elias D C, Nair R R, Mohiuddin T M G, Morozov S V, Blake P, Halsall M P et. al. \emph{Science} 323 610 2009.
\bibitem{8} Samarakoon D K and Wang X Q. \emph{ACS Nano} 12 4017 2009.
\bibitem{9} Leenaerts O, Peelaers H, Hern\'andez-Nieves A D, Partoens B and Peeters F M. \emph{Phys. Rev. B} 82 195436 2010.
\bibitem{10} Bhattacharya A, Bhattacharya S, Majumder C and Das G P. \emph{Phys. Rev. B} 83 033404 2011.
\bibitem{11} Wen X D, Hand L, Labet V, Yang T, Hoffmann R, Ashcroft N W, Oganov A R and Lyakhov O. \emph{Proc. Natl. Acad. Sci. U.S.A.} 108 6833 2011.
\bibitem{12} Samarakoon D K, Chen Z F, Nicolas C and Wang X Q. \emph{Small} 7, 965 2011.
\bibitem{13} Perdew J P and Wang Y. \emph{Phys. Rev. B} 45 13244 1992.
\bibitem{14} Kresse G and Furthm\"{u}ller J. \emph{Phys. Rev. B} 54 11169 1996.
\bibitem{15} Kresse G and Furthm\"{u}ller J. \emph{Comput. Mater. Sci} 6 15 1996.
\bibitem{16} Bl\"{o}chl P E. \emph{Phys. Rev. B} 50 17953 1994.
\bibitem{17} Kresse G and Joubert D. \emph{Phys. Rev. B} 59 1758 1999.
\end{thebibliography}
\end{document}